\documentclass[11pt]{article}
\usepackage{amsmath}
\usepackage{amssymb}

\begin{document}

\title{BPS states in type IIB SUGRA with $SO\left(4\right)\times SO\left(2\right)_{gauged}$
symmetry}

\author{Aristomenis Donos\\
 Physics Department\\
 Brown University\\
 Providence, Rhode Island 02912, USA}

\maketitle
\begin{abstract}
We present an extension of our construction (hep-th/0606199) exhibiting
$SO\left(4\right)\times SO\left(2\right)$ symmetry. We extend the
previously presented ansatz by introducing a $U\left(1\right)$ gauge
field. The presence of the gauge field allows for more general values
of the Killing spinor $U\left(1\right)$ charge. One more time we
identify a four dimensional Kahler structure and a Monge-Ampere type
of equation parametrized by the $U\left(1\right)$ Killing spinor
charge. In addition we identify 2 scalars that parametrize the supersymmetric
solutions, one of which is the Kahler potential.\newpage{} 
\end{abstract}

\section{Introduction}

Supersymmetry provides us with a powerful tool in obtaining or at
least describing solutions of supergravity theories. Using the G-structure
analysis, originally developed in \cite{Gauntlett:2002sc,Gauntlett:2002nw,Gutowski:2003rg,Gauntlett:2004zh,Gauntlett:2004yd},
one is able to write down the constraints that the bosonic fields
need to satisfy in order for the background they create to be supersymmetric.
In general the constraints are general and illuminating but the method
is more fruitful when one makes a sensible reduction based on symmetry
grounds. The method has been applied to several interesting configurations
\cite{Kim:2005ez,Kim:2006qu,Liu:2004hy,Liu:2004ru,Liu:2006is,Lunin:2006xr}
and interesting results were obtained.

One of the most interesting cases was presented in \cite{Lin:2004nb}
where the authors, among other results, demonstrated a one to one
mapping between $1/2$ BPS states in minimal type IIB supergravity
and states in $\mathcal{N}=4$ SYM preserving the same amount of supersymmetry.
The procedure exploits the fact that on both sides the states have
the same moduli space which is parametrized by the phase space of
$N$ non-relativistic massless fermions in a simple harmonic potential.
The field theory study was carried out in \cite{Corley:2001zk} and
\cite{Berenstein:2004kk}. The symplectic form of the moduli space
variables was later computed \cite{Grant:2005qc,Maoz:2005nk} and
was shown to agree with the symplectic form of the matrix model relevant
to the field theory states \cite{Iso:1992ca}.

In this paper we present an obvious extension of our previous effort
\cite{Donos:2006iy} in generalizing the analysis of \cite{Lin:2004nb}
to bosonic states preserving only $SO\left(4\right)\times SO\left(2\right).$
The new element that we are concerned with is the vector field that
can be added to gauge the original $SO\left(2\right).$ Having a non-zero
field strength allows us to have more general $U\left(1\right)$ spinor
charge, which is not constrained by the $SO\left(4\right)$ chirality
of the Killing spinor. As in the ungauged case we are able to demonstrate
a four dimensional Kahler structure. In addition we are also able
to show that the gauge vector can be parametrized by a single scalar
function. We finally show that the supersymmetry constraints give
us a Monge-Ampere type of equation along with a non-linear constrain.
The Bianchi identities that we want the five form to satisfy give
us an additional constrain.

The paper is structured in three sections. In the first section we
present our ansatz and we also show the technical highlights of the
supersymmetry analysis. In the second section we embed known supersymmetric
solutions in our general ansatz where the significance of the $U\left(1\right)$
spinor charge becomes more transparent. In the last section we present
a summary and conclusions. We also include an appendix where we give
the technical details of the supersymmetry analysis for the interested
reader.\newpage{}

\section{The $SO\left(4\right)\times SO\left(2\right)$ symmetric ansatz and
the SUSY analysis}

In this section we will briefly describe the main steps of the supersymmetry
analysis. Following the LLM analysis \cite{Lin:2004nb} we first reduce
the ten dimensional theory by imposing $SO\left(4\right)\times SO\left(2\right)$
symmetry on the fields of minimal type IIB supergravity, namely the
metric $g_{MN}$ and the self-dual five form field strength $F_{M_{1}M_{2}M_{3}M_{4}M_{5}}$.
Our starting point is the ansatz\begin{align}
ds^{2} & =g_{\mu\nu}dx^{\mu}dx^{\nu}+e^{H+G}d\hat{\Omega}_{3}^{2}+e^{H-G}\left(d\psi+A\right)^{2}\nonumber \\
F_{\left(5\right)} & =\hat{F}_{\left(2\right)}\wedge d\hat{\Omega}_{3}+\tilde{F}_{\left(4\right)}\wedge\left(d\psi+A\right).\label{anz}\end{align}
 Where the Greek indices $\mu,\nu=1\ldots6$.

In general, the constraints obtained by the G-structure analysis don't
have to neceserily satisfy the field equations of type IIB supergravity.
The check that guarantees the compatibility of the configuration with
the type IIB field equations is the Bianchi identities that the five
form should satisfy \cite{Gran:2005ct}. This argument is based on
the integrability of the Killing spinor equation. For our case this
means that the various form field strengths that come from the reduction
of the five form have to satisfy\[
\hat{F}_{2}=2\, e^{2G}e^{H}\star_{6}\tilde{F}_{4}\]
 and the Bianchi identity for the five form gives\begin{align}
d\hat{F}_{\left(2\right)} & =0\nonumber \\
d\tilde{F}_{\left(4\right)} & =0\nonumber \\
\tilde{F}_{\left(4\right)}\wedge\mathcal{F} & =0.\label{F_4_W_constrain}\end{align}

The problem that we would like to confront is to identify all the
constraints imposed on the previously bosonic fields so that the Killing
spinor equation\begin{equation}
\mathcal{D_{M}}\eta=\nabla_{M}\eta+\frac{\imath}{480}\Gamma^{M_{1}\ldots M_{5}}F_{M_{1}\ldots M_{5}}\Gamma_{M}\eta=0.\label{killingequation}\end{equation}
 will admit at least one non-trivial solution. After reducing on $S^{3}\times S^{1}$,
as we describe in Appendix A we are left with a six dimensional spinor
$\varepsilon$, a differential equation in six dimensions and two
algebraic ones coming from the reduction on $S^{3}$ and $S^{1}$
respectively\begin{align}
\left[\nabla_{\mu}-\frac{\imath n}{2}A_{\mu}+\frac{1}{4}e^{\frac{1}{2}\left(H-G\right)}\gamma_{7}\gamma^{\nu}\mathcal{F}_{\nu\mu}-\imath N\gamma_{\mu}\right]\varepsilon & =0\label{kileq2}\\
\left[-\alpha e^{-\frac{1}{2}\left(H+G\right)}\gamma_{7}+\imath ne^{-\frac{1}{2}\left(H-G\right)}-\frac{1}{4}e^{\frac{1}{2}\left(H+G\right)}\not\mathcal{F}+\gamma_{7}\gamma^{\lambda}\partial_{\lambda}H\right]\varepsilon & =0\label{kil1}\\
\left[\imath\alpha e^{-\frac{1}{2}\left(H+G\right)}-ne^{-\frac{1}{2}\left(H-G\right)}\gamma_{7}-\imath\gamma^{\lambda}\partial_{\lambda}G-\frac{\imath}{4}e^{\frac{1}{2}\left(H+G\right)}\not\mathcal{F}\gamma_{7}+4N\right]\varepsilon & =0\label{kil2}\\
N=-\frac{1}{4}\not\hat{F}e^{-\frac{3}{2}\left(G+H\right)}\end{align}
 where $n$ is the $U\left(1\right)$ spinor charge and $\alpha$
is the $SO\left(4\right)$ chirality.

At this point we introduce the spinor bilinears that one can construct
from the six dimensional Killing spinor\begin{align}
f_{1} & =\bar{\varepsilon}\gamma_{7}\varepsilon\label{f1}\\
f_{2} & =\imath\bar{\varepsilon}\varepsilon\label{f2}\\
K_{\mu} & =\bar{\varepsilon}\gamma_{\mu}\varepsilon\label{km}\\
L_{\mu} & =\bar{\varepsilon}\gamma_{\mu}\gamma_{7}\varepsilon\label{lm}\\
Y_{\mu\lambda} & =\imath\bar{\varepsilon}\gamma_{\mu\nu}\gamma_{7}\varepsilon\label{yml}\\
V_{\mu\nu} & =\bar{\varepsilon}\gamma_{\mu\nu}\varepsilon\label{vmn}\\
\Omega_{\mu\nu\lambda} & =\imath\bar{\varepsilon}\gamma_{\mu\nu\lambda}\varepsilon.\label{omegamnl}\end{align}
 As we show in Appendix B one can prove that\[
\nabla_{\left(\mu\right.}K_{\left.\nu\right)}=0\]
 which suggests that $K_{\mu}$ is a Killing vector for the six dimensional
metric. At this point we impose the condition\begin{equation}
K^{\mu}\mathcal{F}_{\mu\nu}=0\label{condition}\end{equation}
 which will greatly simplify our analysis. One can then use the Killing
spinor equation and the two projectors to show that\begin{align*}
f_{2} & =\kappa e^{\frac{1}{2}\left(H+G\right)}\\
f_{1} & =\lambda e^{\frac{1}{2}\left(H-G\right)}\end{align*}
 where $\kappa$ and $\lambda$ are integration constants which give
the same form for these bilinears similar to the ones we found in
\cite{Donos:2006iy}. It is also notable that another consequence
of \eqref{condition} is the fact that $L$ is again a closed form
which we can show, by using \eqref{kil1}, that\[
\partial_{\mu}e^{H}=-\frac{\alpha}{\lambda}L_{\mu}.\]
 From the differential equation that $L$ satisfies \eqref{dlm} we
obtain the constrain\[
\mathcal{F}_{\lambda\left[\mu\right.}V_{\:\:\left.\nu\right]}^{\lambda}=0\]
 which will allow us to parametrize the field strength $\mathcal{F}$
by a single scalar function. Using the six dimensional Fierz identities
and the algebraic equation \eqref{kil1} can be used to show that
the two vectors $K$ and $L$ are orthogonal and they have opposite
norms\begin{align*}
L^{2} & =-K^{2}=h^{-2}=f_{1}^{2}+f_{2}^{2}\\
L\cdot K & =0\end{align*}
 and also that\[
L^{\mu}\mathcal{F}_{\mu\nu}=0.\]
 Following similar arguments that where presented in similar configurations
\cite{Liu:2004hy,Lin:2004nb,Donos:2006iy} one can reduce the form
of the metric to\begin{align*}
ds^{2} & =-\frac{1}{h^{2}}\left(dt+C\right)^{2}+h^{2}dy^{2}+e^{-G-H}h_{mn}dx^{m}dx^{n}+ye^{G}d\hat{\Omega}_{3}^{2}+ye^{-G}d\left(\psi+A\right)^{2},\quad m,n=1,\ldots,4\\
L & =-\alpha\, dy.\end{align*}
 At this point we have set $\kappa=\lambda=1$ and $\gamma=-\alpha.$
Where the four dimensional metric $h_{mn}$ also depends on the coordinate
$y$ as a parameter. The three form \eqref{omegamnl} which we show
that can be written as\[
\Omega=e^{-G-H}K\wedge\mathcal{J}\]
 provides us with a complex structure $\mathcal{J}$ for the four
dimensional base space that appears in the previous form of the metric.
The $y$ derivative of the Kahler form is given by a combination a
the field strengths of the vectors $C$ and $A$\[
\frac{-\alpha}{y}\partial_{y}\mathcal{J}=\tilde{d}C-\mathcal{F}\]
 while the $y$ derivative of the vector $C$ is given by\[
\partial_{y}C=-\frac{1}{y}\mathcal{J}\cdot\tilde{d}Z\]
 where the exterior derivative $\tilde{d}$ acts on the coordinates
of the four dimensional space.

The four form can be shown that is given by\begin{align*}
4f_{1}e^{-\frac{1}{2}\left(H-G\right)}\tilde{F}_{\left(4\right)} & =-\frac{2f_{1}^{2}}{f_{1}^{2}+f_{2}^{2}}\, dG\wedge K\wedge I+f_{1}e^{-\frac{1}{2}\left(G+H\right)}\, L\wedge K\wedge dC-\left(f_{2}^{2}+f_{1}^{2}\right)\: dC\wedge I\\
 & \quad+\lambda e^{H-G}I\wedge\mathcal{F}-e^{\frac{1}{2}\left(H-G\right)}\frac{f_{1}^{2}}{f_{2}}\frac{1}{f_{1}^{2}+f_{2}^{2}}L\wedge K\wedge\mathcal{F}\end{align*}
 and the closure of this four form forces the constrain\[
\mathcal{F}\wedge d\left(e^{\frac{1}{2}\left(H-G\right)}Y\right)=0\]
 which can be shown to equivalent to the constrain \eqref{F_4_W_constrain}.

The higher dimensional analog of the LLM Laplace equation is not altered
in our case. As in the ungauged case, we find that the connection
between the Kahler form and the scalar\[
Z=\frac{1}{2}\tanh G\]
 is given by \[
y\partial_{y}\left(\frac{\partial_{y}\mathcal{J}}{y}\right)+\tilde{d}\left(\mathcal{J}\cdot\tilde{d}Z\right)=0.\]

Supersymmetry demands that the volume of the four dimensional satisfies
the differential equation\begin{align*}
\partial_{y}\ln\det\mathcal{G} & =2\frac{e^{-G}}{e^{G}+e^{-G}}\partial_{y}G+2\alpha\frac{n+2\alpha}{y\left(1+e^{2G}\right)}-\frac{2\alpha}{y}\left(n+\alpha\right)\end{align*}
 which is a more general equation than what we had presented in \cite{Donos:2006iy}.
Apart from the $y$ derivative of the volume we would also like to
consider the derivatives with respect to the coordinates of the four
dimensional base space. For this reason we study the Ricci form of
the four dimensional space. The quantity that will help us calculate
the curvature of the Kahler manifold is the 2-form bilinear \[
\mathcal{V}_{\mu\nu}=\varepsilon^{T}\gamma_{\mu\nu}\gamma_{7}\varepsilon.\]
 This 2-form can be shown to be holomorphic and as in other interesting
cases \cite{Kim:2005ez,Kim:2006qu} we can use the differential Killing
spinor equation to calculate the curvature of the Kahler base space.
In the end of the calculation we show that the Ricci form is given
by \eqref{Ricci_Form}.

The volume that correctly reproduces both equations under consideration
is\begin{align*}
\ln\left|\begin{array}{cc}
\partial_{z}\partial_{\bar{z}}K & \partial_{w}\partial_{\bar{z}}K\\
\partial_{z}\partial_{\bar{w}}K & \partial_{w}\partial_{\bar{w}}K\end{array}\right| & =\ln\left(-y\partial_{y}\left(\frac{\partial_{y}K}{y}\right)+yF^{\prime}\left(y\right)\right)+\frac{2\alpha}{y}\left(2\alpha+n\right)\partial_{y}K+4\alpha\left(\alpha+n\right)\Phi\\
 & \qquad-2\alpha\left(n+\alpha\right)\ln y-2\alpha\left(2\alpha+n\right)F\left(y\right)+2\alpha\left(2\alpha+n\right)\ln y\end{align*}
 where the function $F\left(y\right)$ is such that\[
Z+\frac{1}{2}=-y\partial_{y}\left(\frac{\partial_{y}K}{y}\right)+yF^{\prime}\left(y\right).\]
 As we will see for the space-times which are asymptotically $AdS_{5}\times S^{5}$
this function is irrelevant. and the scalar $\Phi$ is introduced
through the field strength\[
\mathcal{F}=\tilde{d}\left(\mathcal{J}\cdot\tilde{d}\Phi\right)\]
 which solves the constrain \eqref{WVconstrain}.

The scalar $\Phi$ can be thought of as being sourced by the constrain\[
\mathcal{F}-\frac{1}{f_{1}^{2}+f_{2}^{2}}i_{L}i_{K}\star_{6}\mathcal{F}=-2e^{-2H}\left(n+\alpha\right)\mathcal{J}\]
 which, generalizes the constrain $n+\alpha=0$ that appears in similar
studies \cite{Liu:2004hy,Lin:2004nb,Donos:2006iy}. As we can see
from the previous constrain, since both the field strength $\mathcal{F}$
and the Kahler form $\mathcal{J}$ are closed with respect to the
four dimensional external differentiation operator $\tilde{d}$ ,
we essentially have that\[
\tilde{d}\star_{4}\mathcal{F}=0\]
 which is compatible with the type IIB field equations.

The two constraints that originate from the ten dimensional Bianchi
identities are simultaneously equivalent to the constrain \[
\mathcal{F}\wedge\mathcal{F}=2\left(n+2\alpha\right)\left(n+\alpha\right)\, y^{-4}\,\mathcal{J}\wedge\mathcal{J}\]
 which ensures that the geometry solves the ten dimensional type IIB
equations of motion.

The last thing that needs to be checked is the closure of the two
form $\hat{F}_{\left(2\right)}$. Using the duality relation \eqref{Fduality}
we can determine the two form $\hat{F}_{\left(2\right)}$ whose components
read\begin{align*}
\hat{F}_{tx} & =-\frac{1}{2}e^{\frac{3}{2}\left(G+H\right)}f_{2}\partial_{x}G\\
\hat{F}_{ty} & =-\frac{1}{4}\partial_{y}e^{2\left(G+H\right)}\\
\hat{F}_{yx} & =\frac{1}{4}C_{x}\partial_{y}e^{2\left(G+H\right)}-\frac{1}{8}e^{G+H}\left(e^{G}+e^{-G}\right)\,\mathcal{J}_{x}^{\; x_{1}}\partial_{x^{1}}z\\
\hat{F}_{x_{1}x_{2}} & =\frac{1}{2}\mathcal{J}_{x_{1}x_{2}}-\frac{1}{4}e^{H}\partial_{y}\mathcal{J}_{x_{1}x_{2}}-\frac{1}{4}e^{2G+H}\partial_{y}\mathcal{J}_{x_{1}x_{2}}-\frac{1}{4}e^{2\left(G+H\right)}\mathcal{F}_{x_{1}x_{2}}-\frac{1}{4}\partial_{\left[x_{1}\right.}e^{2\left(G+H\right)}C_{\left.x_{2}\right]}.\end{align*}
 Checking the closure of the two form amounts to using the differential
constraints that we have already identified.

\section{Three examples}

In this section we will reduce our general configuration to some known
examples. This will help us illustrate the physical meaning of the
$U\left(1\right)$ charge $n$ that enters in our analysis.

\subsection*{Ungauged $SO\left(2\right)$ ($n=-\alpha$)}

The first example that we present will reduce the more general ansatz
\eqref{anz} to the ungauged case we had originally considered in
\cite{Donos:2006iy}. We take $n=-\alpha$ and from the constrain
\eqref{WW_equals_JJ} we see that the field strength of the gauge
vector has to satisfy\[
\mathcal{F}\wedge\mathcal{F}=0.\]
 From our second constrain of the field strength \eqref{w_constrain}
we also have that\[
\mathcal{F}=\star_{4}\mathcal{F}\]
 and because these forms carry only the four dimensional Euclidean
indices we are forced to\[
\mathcal{F}=0.\]
 Finally, we see that the gauge field is a pure gauge which we can
eliminate by shifting $\psi$. In this case we see that we recover
the case of the ungauged solutions. As one can easily check the our
more general equation \eqref{mong_Amp} reduces to the equation we
had proposed previously proposed in \cite{Donos:2006iy}.

\subsection*{LLM ($n=-2\alpha$)}

The class of LLM solutions \cite{Lin:2004nb} should be present in
our previous case. However, writing down the correct Kahler potential
and making the correct identification of coordinates looks like a
rather non-trivial task. What we would like is to exploit is the presence
of the gauge field. \[
\mathcal{F}\wedge\mathcal{F}=0\]
 but $\mathcal{F}$ is non-zero now as we can see from the constraints
that it satisfies. We split the four dimensional space in pairs $\left(x_{1},x_{2}\right)$
and $\left(x_{3},x_{4}\right)$. We now write\begin{align*}
\mathcal{F} & =A_{3}\left(x_{3},x_{4}\right)dx^{3}+A_{4}\left(x_{3},x_{4}\right)dx^{4}\\
C & =C_{1}\left(x_{1},x_{2},y\right)dx^{1}+C_{2}\left(x_{1},x_{2},y\right)dx^{2}\\
\mathcal{J} & =\mathcal{J}^{\left(1\right)}+y^{2}\mathcal{J}^{\left(2\right)}\end{align*}
 and the metric takes the form\[
ds^{2}=-\frac{1}{h^{2}}\left(dt+C\right)^{2}+h^{2}dy^{2}+\frac{1}{ye^{G}}ds_{1}^{2}+ye^{G}d\hat{\Omega}_{3}^{2}+ye^{-G}\left[ds_{2}^{2}+\frac{1}{4}d\left(\tilde{\psi}+2A\right)^{2}\right].\]
 Using equations \eqref{Ricci_Form} and \eqref{dyJ_dC_W} we have
that for the two dimensional space spanned by $\left(x_{3},x_{4}\right)$
the Ricci tensor and the vector field $A$ satisfy\begin{align*}
R^{\left(2\right)} & =4g^{\left(2\right)}\\
dA & =2\mathcal{J}^{\left(2\right)}.\end{align*}
 From the first equation we see that the second two dimensional space
is locally a two-sphere of radius $\frac{1}{2}$ for which we may
write\[
\mathcal{J}^{\left(2\right)}=\frac{1}{4}\sigma_{1}\wedge\sigma_{2}\]
 and from the second equation that we had for the vector field we
see that\[
d\left(\tilde{\psi}+2A\right)=d\sigma_{3}\]
 where $\sigma_{i}$ are the left invariant one forms on $S^{3}.$
From equation \eqref{mong_Amp} we have that\[
ds_{1}^{2}=\left(Z+\frac{1}{2}\right)\left[dx_{1}^{2}+dx_{2}^{2}\right]\]
 and as we can see from \eqref{LLM_type} the function $Z$ must satisfy\[
\left(\partial_{1}^{2}+\partial_{2}^{2}\right)Z+y\partial_{y}\left(\frac{\partial_{y}Z}{y}\right)=0.\]
 Unlike the ungauged case $n=-\alpha$ where even the very symmetric
$AdS_{5}\times S^{5}$ looks very non-trivial, we see that after the
inclusion of the $U\left(1\right)$ fiber, the whole class of LLM
solutions comes out very naturally.

\subsection*{$\mathbf{AdS_{5}\times}$Sasaki-Einstein ($n=-3\alpha$)}

In this case we have that\[
\mathcal{F}\wedge\mathcal{F}=4y^{-4}\mathcal{J}\wedge\mathcal{J}\]
 and we write the metric as\[
ds^{2}=-\frac{1}{h^{2}}{dt}^{2}+h^{2}dy^{2}+ye^{G}d\hat{\Omega}_{3}^{2}+ye^{-G}\left[ds_{4}^{2}\left(x^{i}\right)+d\left(\tilde{\psi}+A\right)^{2}\right]\]
 where we have rescaled pulled out a factor $y^{2}$ from the four
dimensional metric. In this case we set $e^{G}=y$ and\[
Z=\frac{1}{2}\frac{1-y^{2}}{1+y^{2}}.\]
 From the Ricci form of the four dimensional space \eqref{Ricci_Form}
and equation \eqref{dyJ_dC_W} we can see that the four dimensional
Kahler manifold becomes Einstein-Kahler\[
\mathcal{R}=6\mathcal{J}.\]
 On the other hand the Kahler form is related to the field strength
$\mathcal{F}$ \[
\mathcal{F}=2\mathcal{J}.\]
 The above conditions guarantee that the 5 dimensional spanned by
the coordinates $\left(\tilde{\psi},x_{1},\ldots,x_{4}\right)$ is
a five dimensional Sasaki-Einstein space. The final form of the metric
is\[
ds^{2}=-\left(y^{2}+1\right)dt^{2}+\frac{dy^{2}}{y^{2}+1}+y^{2}d\hat{\Omega}_{3}^{2}+ds_{4}^{2}\left(x^{i}\right)+d\left(\tilde{\psi}+A\right)^{2}.\]

\section{Conclusions and Summary}

We have enlarged our previous $SO\left(4\right)\times SO\left(2\right)$
symmetric ansatz \cite{Donos:2006iy} to include a vector field which
gauges the $SO\left(2\right)$ symmetry. The feature that persists
after the inclusion of the new field is the four dimensional Kahler
structure that we identified. This might not be very surprising since
a six dimensional Kahler structure has been nicely identified in \cite{Kim:2005ez}
for the $SO\left(4\right)$ symmetric case. In addition to the Kahler
potential a new scalar makes its appearance through the gauge field
as someone would expect. The two scalars are coupled through constraints
which makes them dependent.

The analysis we have presented in this paper is in no sense complete
and more work is required. The configurations we have analyzed are
all BPS but the constraints that one needs to solve are non-linear.
It would be interesting to show at least the integrability of the
above constraints and give a sense of superposition.

It would be very interesting to discover in this less supersymmetric
setup a higher dimensional fermion droplet, that would parametrize
regular solutions, and make contact with a multi-matrix model picture
analysis \cite{deMelloKoch:2003pv,Donos:2005vm,Berenstein:2005aa,Lin:2006tr,Biswas:2006tj,Vazquez:2006hd}.
In this context it would be interesting to think of those geometries
as a back reaction to Mikhailov's giant gravitons \cite{Mikhailov:2000ya}.

\section*{Acknowledgments}

I would like to thank the MCTP high energy theory group for the hospitality
during the beginning of this work. I would especially like to thank
Sera Cremonini, Hai Lin, James Liu and Diana Vaman for many useful
discussions, comments and their collaboration. Finally, I thank Antal
Jevicki and Robert McNees for many invaluable discussions.

\newpage{}

\appendix

\section{Killing spinor equation reduction}

We start this appendix by giving the reduction of the spin connection
on $S^{3}\times S^{1}$\begin{align*}
\hat{\omega}_{cd} & =\omega_{cd}-\frac{1}{2}e^{H-G}\mathcal{F}_{cd}A-\frac{1}{2}e^{H-G}\mathcal{F}_{cd}d\psi\\
\hat{\omega}_{d\hat{m}} & =-\frac{1}{2}e^{\frac{1}{2}\left(H+G\right)}\hat{e}_{\hat{m}\hat{\mu}}\partial_{d}\left(H+G\right)dx^{\hat{\mu}}\\
\hat{\omega}_{10d} & =\frac{1}{2}e^{\left(H-G\right)}\mathcal{F}_{d\mu}dx^{\mu}+\frac{1}{2}e^{\frac{1}{2}\left(H-G\right)}\partial_{d}\left(H-G\right)A+\frac{1}{2}e^{\frac{1}{2}\left(H-G\right)}\partial_{d}\left(H-G\right)d\psi\\
\mathcal{F} & =dA\end{align*}
 where $c,d=1\ldots6$ are the tangent indices of the dimensional
space, $\hat{m}=1\ldots3$ are the tangent indices of $S^{3}$.and
$\omega_{cd}$ is the spin connection of the six dimensional space.

We decompose the gamma matrices $\Gamma_{M}$in the following way\begin{align*}
\Gamma_{\mu} & =\gamma_{\mu}\otimes\hat{\sigma}_{1}\otimes\mathbb{I}_{2}\\
\Gamma_{\hat{\mu}} & =\mathbb{I}_{8}\otimes\hat{\sigma}_{2}\otimes\sigma_{\hat{\mu}}\\
\Gamma_{10} & =\gamma_{7}\otimes\hat{\sigma}_{1}\otimes\mathbb{I}_{2}\end{align*}
 where $\gamma_{7}=\gamma_{1}\ldots\gamma_{6}$ , $\Gamma_{11}=\Gamma_{1}\ldots\Gamma_{10}=\mathbb{I}_{8}\otimes\hat{\sigma}_{3}\otimes\mathbb{I}_{2}$
and the chirality condition for the IIB spinors gives\[
\Gamma_{11}\eta=\eta\Rightarrow\hat{\sigma}_{3}\eta=\eta.\]
 For the spinors we give the decomposition\[
\eta=\epsilon\otimes\left[\begin{array}{c}
1\\
0\end{array}\right]\otimes\chi_{\alpha}=\varepsilon\otimes\chi_{\alpha}\]
 where\[
\hat{\nabla}_{\hat{a}}\chi_{\alpha}=\frac{i\alpha}{2}\sigma_{\hat{a}}\chi_{\alpha},\qquad a=\pm1.\]
 For dependence on the coordinates we have\[
\varepsilon\left(x^{\mu},\hat{\Omega}_{3},\psi\right)=e^{\frac{\imath}{2}n\psi}\varepsilon\left(x^{\mu},\hat{\Omega}_{3}\right)\]
 which gives\[
\imath\partial_{\psi}\varepsilon=-\frac{n}{2}\varepsilon.\]
 For the covariant derivatives we have\begin{align*}
\hat{\nabla}_{\mu} & =\nabla_{\mu}-\frac{1}{8}e^{H-G}\not\mathcal{F}A_{\mu}+\frac{1}{4}e^{\frac{1}{2}\left(H-G\right)}\gamma_{7}\gamma^{\nu}\mathcal{F}_{\nu\mu}+\frac{1}{4}e^{\frac{1}{2}\left(H-G\right)}\gamma_{7}\gamma^{\nu}\partial_{\nu}\left(H-G\right)A_{\mu}\\
\nabla_{\hat{\mu}} & =\hat{\nabla}_{\hat{\mu}}-\frac{\imath}{4}\sigma_{\hat{\mu}}\hat{\sigma}_{3}\gamma^{\nu}\partial_{\nu}\left(G+H\right)\\
\nabla_{\psi} & =\partial_{\psi}-\frac{1}{8}e^{H-G}\not\mathcal{F}+\frac{1}{4}\gamma_{7}\gamma^{\mu}\partial_{\mu}\left(H-G\right)\end{align*}
 and for the Dirac matrices we have that\[
\Gamma_{\mu}=\gamma_{\mu}\hat{\sigma}_{1}+e^{\frac{1}{2}\left(H-G\right)}A_{\mu}\gamma_{7}\hat{\sigma}_{1}.\]
 The ten dimensional Killing spinor equation reads\begin{equation}
\nabla_{M}\eta+\frac{\imath}{480}\Gamma^{M_{1}\ldots M_{5}}F_{M_{1}\ldots M_{5}}\Gamma_{M}\eta=0.\end{equation}
 For the second term we have\begin{align*}
M & =\frac{\imath}{480}\Gamma^{M_{1}\ldots M_{5}}F_{M_{1}\ldots M_{5}}\\
 & =\frac{\imath}{480}\left[10\Gamma^{M_{1}M_{2}}\hat{F}_{M_{1}M_{2}}e^{-\frac{3}{2}\left(G+H\right)}\Gamma^{\hat{a}\hat{b}\hat{c}}\varepsilon_{\hat{a}\hat{b}\hat{c}}+5\Gamma^{M_{1}M_{2}M_{3}M_{4}}\Gamma^{\psi}e^{-\frac{1}{2}\left(H-G\right)}\tilde{F}_{M_{1}M_{2}M_{3}M_{4}}\right]\end{align*}
 after using the two duality relations\begin{align*}
\Gamma^{M_{1}\ldots M_{5}} & =-\frac{1}{6!}\varepsilon^{M_{1}\ldots M_{11}}\Gamma_{M_{6}\ldots M_{11}}\end{align*}
 and\begin{equation}
\tilde{F}_{M_{1}\ldots M_{4}}=-\frac{1}{2}e^{-2G-H}\varepsilon_{M_{1}\ldots M_{6}}\hat{F}^{M^{5}M^{6}}\label{Fduality}\end{equation}
 we have\begin{align*}
M & =-\frac{1}{8}\gamma^{\mu_{1}\mu_{2}}\hat{F}_{\mu_{1}\mu_{2}}e^{-\frac{3}{2}\left(G+H\right)}\hat{\sigma}_{2}\left(1+\Gamma_{11}\right)\\
 & =-\frac{1}{8}\not\hat{F}\hat{\sigma}_{2}\left(1+\hat{\sigma}_{3}\right).\end{align*}
 The reduced equation now gives one diferrential and two algebraic
constrains for the Killing spinor\begin{align}
\left[\nabla_{\mu}-\frac{1}{8}e^{H-G}\not\mathcal{F}A_{\mu}+\frac{1}{4}e^{\frac{1}{2}\left(H-G\right)}\gamma_{7}\gamma^{\nu}W_{\nu\mu}\right.\label{kileq}\\
\left.+\frac{1}{4}e^{\frac{1}{2}\left(H-G\right)}\gamma_{7}\gamma^{\nu}\partial_{\nu}\left(H-G\right)A_{\mu}-\imath N\gamma_{\mu}-\imath e^{\frac{1}{2}\left(H-G\right)}N\gamma_{7}A_{\mu}\right]\varepsilon & =0\nonumber \\
\left[\frac{\imath\alpha}{2}e^{-\frac{1}{2}\left(H+G\right)}-\frac{i}{4}\gamma^{\lambda}\partial_{\lambda}\left(H+G\right)+N\right]\varepsilon & =0\label{kilS3}\\
\left[\frac{\imath n}{2}e^{-\frac{1}{2}\left(H-G\right)}-\frac{1}{8}e^{\frac{1}{2}\left(H-G\right)}\not\mathcal{F}+\frac{1}{4}\gamma_{7}\gamma^{\lambda}\partial_{\lambda}\left(H-G\right)-\imath\gamma_{7}N\right]\varepsilon & =0\label{kilS1}\end{align}
 where\[
N=-\frac{1}{4}\not\hat{F}e^{-\frac{3}{2}\left(G+H\right)}.\]
 The last equations can be written after linear combinations as\begin{align*}
\left[\nabla_{\mu}-\frac{\imath n}{2}A_{\mu}+\frac{1}{4}e^{\frac{1}{2}\left(H-G\right)}\gamma_{7}\gamma^{\nu}\mathcal{F}_{\nu\mu}-\imath N\gamma_{\mu}\right]\varepsilon & =0\\
\left[-\alpha e^{-\frac{1}{2}\left(H+G\right)}\gamma_{7}+\imath ne^{-\frac{1}{2}\left(H-G\right)}-\frac{1}{4}e^{\frac{1}{2}\left(H+G\right)}\not\mathcal{F}+\gamma_{7}\gamma^{\lambda}\partial_{\lambda}H\right]\varepsilon & =0\\
\left[\imath\alpha e^{-\frac{1}{2}\left(H+G\right)}-ne^{-\frac{1}{2}\left(H-G\right)}\gamma_{7}-\imath\gamma^{\lambda}\partial_{\lambda}G-\frac{\imath}{4}e^{\frac{1}{2}\left(H+G\right)}\not\mathcal{F}\gamma_{7}+4N\right]\varepsilon & =0.\end{align*}

\section{Supersymmetry analysis}

When looking at Fierz identities involving the above bilinears it
will be more convenient to work with the following equivalent bilinears\begin{align}
Z^{+} & =-f_{1}-\imath f_{2}=2\bar{\varepsilon_{+}}\varepsilon_{-}\label{Z+}\\
Z^{-} & =f_{1}-\imath f_{2}=2\bar{\varepsilon_{-}}\varepsilon_{+}\label{Z-}\\
l_{\mu}^{+} & =L_{\mu}+K_{\mu}=2\bar{\varepsilon_{+}}\gamma_{\mu}\varepsilon_{+}\label{l+}\\
l_{\mu}^{-} & =-L_{\mu}+K_{\mu}=2\bar{\varepsilon_{-}}\gamma_{\mu}\varepsilon_{-}\label{l-}\\
q_{\mu\nu\lambda}^{\pm} & =\imath\bar{\varepsilon_{\pm}}\gamma_{\mu\nu\lambda}\varepsilon_{\pm}\label{q+-}\end{align}
 where as usually\begin{align}
\varepsilon_{\pm} & =\frac{1}{2}\left(\mathbb{I}_{8}\pm\gamma_{7}\right)\varepsilon\\
\gamma_{7}\varepsilon_{\pm} & =\pm\varepsilon_{\pm}.\end{align}
 We will also make extensive use of the duality relation in six dimensions
between the gamma matrices\begin{equation}
\gamma^{a_{1}\ldots a_{n}}=\frac{\left(-1\right)^{\left[\frac{n}{2}\right]+1}}{\left(6-n\right)!}\varepsilon^{a_{1}\ldots a_{n}b_{b+1}\ldots b_{6-n}}\gamma_{b_{1}\ldots b_{6-n}}\gamma_{7}\label{gammaduality}\end{equation}
 and the Fierz rearrangement for commuting spinors\begin{align}
\bar{\psi_{1}}\psi_{2}\bar{\psi}_{3}\psi_{4} & =\frac{1}{8}\left[\bar{\psi_{1}}\psi_{4}\bar{\psi}_{3}\psi_{2}+\bar{\psi_{1}}\gamma_{7}\psi_{4}\bar{\psi}_{3}\gamma_{7}\psi_{2}-\frac{1}{2}\bar{\psi_{1}}\gamma_{\mu\nu}\psi_{4}\bar{\psi}_{3}\gamma^{\mu\nu}\psi_{2}-\frac{1}{2}\bar{\psi_{1}}\gamma_{\mu}\gamma_{7}\psi_{4}\bar{\psi}_{3}\gamma^{\mu\nu}\gamma_{7}\psi_{2}\right]\nonumber \\
 & +\frac{1}{8}\left[\bar{\psi_{1}}\gamma_{\mu}\psi_{4}\bar{\psi}_{3}\gamma^{\mu}\psi_{2}-\bar{\psi_{1}}\gamma_{\mu}\gamma_{7}\psi_{4}\bar{\psi}_{3}\gamma^{\mu}\gamma_{7}\psi_{2}\right]\nonumber \\
 & -\frac{1}{96}\left[\bar{\psi_{1}}\gamma_{\mu\nu\lambda}\psi_{4}\bar{\psi}_{3}\gamma^{\mu\nu\lambda}\psi_{2}-\bar{\psi_{1}}\gamma_{\mu\nu\lambda}\gamma_{7}\psi_{4}\bar{\psi}_{3}\gamma^{\mu\nu\lambda}\gamma_{7}\psi_{2}\right].\label{Fierz}\end{align}

\subsection{Diferrential relations}

Taking a covariant derivative of \eqref{f1} and using equation \eqref{kileq2}
we may write\begin{align}
\nabla_{\mu}f_{1} & =\frac{\imath}{4}\bar{\varepsilon}\left(\gamma_{\mu}\gamma_{\kappa\lambda}\gamma_{7}+\gamma_{\kappa\lambda}\gamma_{\mu}\gamma_{7}\right)\varepsilon\, F^{\kappa\lambda}e^{-\frac{3}{2}\left(G+H\right)}-\frac{1}{2}e^{\frac{1}{2}\left(H-G\right)}\bar{\varepsilon}\gamma^{\nu}\varepsilon\mathcal{F}_{\nu\mu}\nonumber \\
 & =\frac{1}{2}\star\left(F\wedge\Omega\right)_{\mu}e^{-\frac{3}{2}\left(G+H\right)}-\frac{1}{2}e^{\frac{1}{2}\left(H-G\right)}K^{\nu}\mathcal{F}_{\nu\mu}\label{df1}\\
 & =-\frac{1}{3!}F_{\mu\rho\sigma\tau}\Omega^{\rho\sigma\tau}e^{\frac{1}{2}\left(G-H\right)}-\frac{1}{2}e^{\frac{1}{2}\left(H-G\right)}K^{\nu}\mathcal{F}_{\nu\mu}\end{align}
 Following the same procedure for \eqref{f2} we have\begin{align}
\nabla_{\mu}f_{2} & =-\frac{1}{2}\bar{\varepsilon}\left(\gamma_{\lambda}g_{\mu\kappa}-g_{\mu\lambda}\gamma_{\kappa}\right)\varepsilon\, F^{\kappa\lambda}e^{-\frac{3}{2}\left(G+H\right)}\nonumber \\
 & =-e^{-\frac{3}{2}\left(G+H\right)}F_{\mu\lambda}K^{\lambda}.\label{df2}\end{align}
 For \eqref{km} we have\begin{align}
\nabla_{\mu}K_{\rho} & =e^{-\frac{3}{2}\left(G+H\right)}f_{2}F_{\mu\rho}-e^{-\frac{3}{2}\left(G+H\right)}\frac{1}{4}F^{\kappa\lambda}\varepsilon_{\mu\rho\pi\tau\kappa\lambda}Y^{\pi\tau}+\frac{1}{2}e^{\frac{1}{2}\left(H-G\right)}f_{1}\mathcal{F}_{\lambda\mu}\label{dkm}\\
 & =e^{-\frac{3}{2}\left(G+H\right)}f_{2}F_{\mu\rho}+\frac{1}{2}e^{\frac{1}{2}\left(G-H\right)}F_{\mu\rho\pi\tau}Y^{\pi\tau}+\frac{1}{2}e^{\frac{1}{2}\left(H-G\right)}f_{1}\mathcal{F}_{\lambda\mu}\end{align}
 For \eqref{lm} we have\begin{align}
\nabla_{\mu}L_{\rho} & =\frac{\imath}{4}F^{\kappa\lambda}e^{-\frac{3}{2}\left(G+H\right)}\bar{\varepsilon}\left(\gamma_{\rho}\gamma_{\kappa\lambda}\gamma_{\mu}+\gamma_{\mu}\gamma_{\kappa\lambda}\gamma_{\rho}\right)\gamma_{7}\varepsilon+\frac{1}{4}e^{\frac{1}{2}\left(H-G\right)}\bar{\varepsilon}\left(\gamma^{\nu}\gamma_{\rho}-\gamma_{\rho}\gamma^{\nu}\right)\varepsilon\mathcal{F}_{\nu\mu}\nonumber \\
 & =e^{-\frac{3}{2}\left(G+H\right)}\left[F_{\mu}^{\:\lambda}Y_{\lambda\rho}+F_{\rho}^{\:\lambda}Y_{\lambda\mu}+\frac{1}{2}g_{\mu\rho}F^{\kappa\lambda}Y_{\kappa\lambda}\right]-\frac{1}{2}e^{\frac{1}{2}\left(H-G\right)}V_{\rho}^{\:\nu}\mathcal{F}_{\nu\mu}\label{dlm}\end{align}
 where we used the duality \eqref{gammaduality}. One more equation
for \eqref{vmn} is given by\begin{equation}
\nabla_{\gamma}V_{\delta\epsilon}=-e^{-\frac{3}{2}\left(G+H\right)}\left[g_{\gamma\left[\epsilon\right.}\Omega_{\left.\delta\right]\alpha\beta}F^{\alpha\beta}-F_{\gamma}^{\:\beta}\Omega_{\beta\delta\epsilon}+2\Omega_{\alpha\gamma\left[\delta\right.}F_{\:\left.\epsilon\right]}^{\alpha}\right]+e^{\frac{1}{2}\left(H-G\right)}L_{\gamma}\mathcal{F}_{\delta\epsilon}.\label{dvmn}\end{equation}

From equation \eqref{dkm} we conclude that\[
\nabla_{\left(\mu\right.}K_{\left.\nu\right)}=0\]
 which suggests that $K_{\mu}$ is a Killing vector.

As we will later see in \eqref{dHL}, a consequence of \eqref{condition},
the 1-form $L_{\mu}$ is the derivative of a scalar and so we have
from \eqref{dlm} the constraint\begin{equation}
dL=0\Rightarrow\mathcal{F}_{\lambda\left[\mu\right.}V_{\:\:\left.\nu\right]}^{\lambda}=0.\label{WVconstrain}\end{equation}

We now take a derivative of \eqref{omegamnl} giving us\begin{align*}
\nabla_{\kappa}\Omega_{\mu\nu\lambda} & =\frac{1}{4}e^{-\frac{3}{2}\left(G+H\right)}F^{\pi\rho}\bar{\varepsilon}\left(\gamma_{\mu\nu\lambda}\gamma_{\pi\rho}\gamma_{\kappa}-\gamma_{\kappa}\gamma_{\pi\rho}\gamma_{\mu\nu\lambda}\right)\varepsilon+\frac{3}{2}e^{\frac{1}{2}\left(H-G\right)}Y_{\left[\mu\nu\right.}\mathcal{F}_{\left.\lambda\right]\kappa}.\end{align*}
 After antisymmetrization we have that \begin{equation}
d\Omega_{\kappa\lambda\mu\nu}=4f_{1}e^{-\frac{1}{2}\left(H-G\right)}\tilde{F}_{\kappa\lambda\mu\nu}+6\, e^{\frac{1}{2}\left(H-G\right)}Y_{\left[\mu\nu\right.}\mathcal{F}_{\left.\lambda\kappa\right]}.\label{dOmega_equals_F}\end{equation}
 For its dual we have the equation\begin{align*}
\nabla_{\kappa}\left(\star\Omega\right)_{\mu\nu\lambda} & =-\frac{1}{4}e^{-\frac{3}{2}\left(G+H\right)}F^{\pi\rho}\bar{\varepsilon}\left(\gamma_{\mu\nu\lambda}\gamma_{\pi\rho}\gamma_{\kappa}+\gamma_{\kappa}\gamma_{\pi\rho}\gamma_{\mu\nu\lambda}\right)\gamma_{7}\varepsilon-\frac{\imath}{2}e^{\frac{1}{2}\left(H-G\right)}\bar{\varepsilon}\gamma_{\mu\nu\lambda}^{\qquad\rho}\varepsilon\mathcal{F}_{\rho\kappa}.\end{align*}
 Antisymmetrizing the last equation in $\kappa,\mu,\nu,\lambda$ we
obtain\begin{equation}
\left(d\star\Omega\right)_{\kappa\mu\nu\lambda}=2\bar{\varepsilon}\left(\gamma_{\mu\nu\lambda\kappa}N+N\gamma_{\kappa\mu\nu\lambda}\right)\gamma_{7}\varepsilon-e^{\frac{1}{2}\left(H-G\right)}\mathcal{F}_{\:\:\left[\kappa\right.}^{\pi}\varepsilon_{\left.\mu\nu\lambda\right]\pi\alpha\beta}Y^{\alpha\beta}\label{d*omega}\end{equation}
 which as we see later promotes a 4-dimensional submanifold to a Kahler
manifold.

\subsubsection*{Algebraic relations}

Using $\bar{\psi}_{1}=\bar{\varepsilon_{+}}\gamma_{\mu}$, $\psi_{2}=\varepsilon_{+}$,
$\bar{\psi_{3}}=\bar{\varepsilon}_{+}$ and $\psi_{4}=\gamma_{\mu}\varepsilon_{+}$
in \eqref{Fierz} we obtain\begin{equation}
l_{\mu}^{+}l^{+\mu}=0.\label{l+l+}\end{equation}
 In a similar manner one can obtain the relation\begin{equation}
l_{\mu}^{-}l^{-\mu}=0.\label{l-l-}\end{equation}
 Using $\bar{\psi}_{1}=\bar{\varepsilon_{+}}$, $\psi_{2}=\varepsilon_{-}$,
$\bar{\psi_{3}}=\bar{\varepsilon}_{-}$ and $\psi_{4}=\varepsilon_{+}$
we obtain\begin{equation}
Z^{+}Z^{-}=\frac{1}{4}l_{\mu}^{+}l^{-\mu}+\frac{1}{48}q^{+\pi\rho\sigma}q_{\pi\rho\sigma}^{-}.\label{ZZ}\end{equation}
 Choosing $\bar{\psi}_{1}=\bar{\varepsilon_{+}}\gamma^{\mu}$, $\psi_{2}=\varepsilon_{+}$,
$\bar{\psi_{3}}=\imath\bar{\varepsilon}_{+}$ and $\psi_{4}=\gamma_{\mu\nu\lambda}\varepsilon_{+}$we
have\begin{equation}
l^{+\mu}q_{\mu\nu\lambda}^{+}=-l^{+\mu}q_{\mu\nu\lambda}^{+}\Rightarrow i_{l^{+}}q^{+}=0.\label{l+q+}\end{equation}
 Using the selfduality\[
\star q^{+}=q^{+}\]
 we have also that\[
l^{+}\wedge\, q^{+}=0.\]
 In a very similar way one can prove that\[
i_{l^{-}}q^{-}=0\]
 and from the anti-selfduality\[
\star q^{-}=-q^{-}\]
 follows that\[
l^{-}\wedge\, q^{-}=0.\]
 The above two relations lead us to the conclusion that in a frame
where the metric can be written as\begin{equation}
ds^{2}=e^{+}e^{-}+\delta_{ab}e^{a}e^{b},\qquad a,b=1\ldots4\label{metric}\end{equation}
 we must have\begin{align*}
l^{+} & =e^{+}\\
l^{-} & =e^{-}\end{align*}
 and for the three forms\begin{align}
q^{+} & =l^{+}\wedge\, I\label{Idef}\\
q^{-} & =l^{-}\wedge\, J\label{Jdef}\end{align}
 where\begin{align}
I & =\frac{1}{2}I_{ab}e^{a}\wedge\, e^{b}\label{Idef2}\\
J & =\frac{1}{2}J_{ab}e^{a}\wedge\, e^{b}.\label{Jdef2}\end{align}
 From the dualities of $q^{\pm}$we conclude that $I$ and J are anti-selfdual
with respect to the four dimensional space. From \eqref{Idef} and
\eqref{Jdef} we conclude that\begin{align}
q^{+\pi\rho\sigma}q_{\pi\rho\sigma}^{-} & =3\left[\left(l^{+}\cdot l^{-}\right)\left(I^{ab}J_{ab}\right)+2l^{+m}l^{-n}I_{nk}J_{\; m}^{k}\right].\nonumber \\
 & =3\left(l^{+}\cdot l^{-}\right)\left(I^{ab}J_{ab}\right).\label{q+q-}\end{align}
 We are now considering \eqref{kilS3}, \eqref{kilS1} and their conjugates\begin{align}
\left[\frac{\imath\alpha}{2}e^{-\frac{1}{2}\left(H+G\right)}-\frac{i}{4}\gamma^{\lambda}\partial_{\lambda}\left(H+G\right)+N\right]\varepsilon & =0\nonumber \\
\bar{\varepsilon}\left[-\frac{\imath\alpha}{2}e^{-\frac{1}{2}\left(H+G\right)}-\frac{\imath}{4}\gamma^{\lambda}\partial_{\lambda}\left(G+H\right)-N\right] & =0\label{ckilS3}\\
\left[\frac{\imath n}{2}e^{-\frac{1}{2}\left(H-G\right)}+\frac{1}{4}\gamma_{7}\gamma^{\lambda}\partial_{\lambda}\left(H-G\right)-\frac{1}{8}e^{\frac{1}{2}\left(H-G\right)}\not\mathcal{F}-\imath\gamma_{7}N\right]\varepsilon & =0\nonumber \\
\bar{\varepsilon}\left[-\frac{\imath n}{2}e^{-\frac{1}{2}\left(H-G\right)}+\frac{1}{4}\gamma^{\lambda}\gamma_{7}\partial_{\lambda}\left(H-G\right)+\frac{1}{8}e^{\frac{1}{2}\left(H-G\right)}\not\mathcal{F}+\imath\gamma_{7}N\right] & =0.\label{ckilS1}\end{align}

\subsubsection*{Vector identities}

At this point it is useful to see that\[
\left[\gamma_{\mu},N\right]=-\frac{1}{2}e^{-\frac{3}{2}\left(G+H\right)}F_{\mu}^{\;\lambda}\gamma_{\lambda}.\]
 Multiplying \eqref{kilS3} by $\bar{\varepsilon}\gamma_{\mu}$, \eqref{ckilS3}
by $\gamma_{\mu}\varepsilon$ and adding the two equations we have\begin{align}
\bar{\varepsilon}\left[\frac{\imath}{4}\left\{ \gamma_{\mu},\gamma^{\lambda}\right\} \partial_{\lambda}\left(H+G\right)+\left[\gamma_{\mu},N\right]\right]\varepsilon & =0\Rightarrow\nonumber \\
f_{2}\partial_{\mu}\left(H+G\right)-e^{-\frac{3}{2}\left(G+H\right)}F_{\mu}^{\:\lambda}K_{\lambda} & =0\end{align}
 which in combination with \eqref{df2} gives the equation\begin{align}
\partial_{\mu}f_{2} & =\frac{1}{2}f_{2}\partial_{\mu}\left(H+G\right)\Rightarrow\label{df22}\\
f_{2} & =\kappa e^{\frac{1}{2}\left(H+G\right)}.\nonumber \end{align}
 We now turn to \eqref{kilS1} and \eqref{ckilS1} , we multiply the
first by $\bar{\varepsilon}\gamma_{\mu}$, the second one by $\gamma_{\mu}\varepsilon$
we add them and in combination with \eqref{condition} yields\begin{align*}
\bar{\varepsilon}\left[-\frac{1}{4}\gamma_{7}\left\{ \gamma_{\mu},\gamma^{\lambda}\right\} \partial_{\lambda}\left(H-G\right)-\imath\left\{ \gamma_{\mu},N\right\} \gamma_{7}\right]\varepsilon & =0\\
-f_{1}\partial_{\mu}\left(H-G\right)+\star\left(F\wedge\Omega\right)_{\mu}e^{-\frac{3}{2}\left(G+H\right)} & =0\end{align*}
 which in combination with \eqref{df1} yields\begin{align}
\partial_{\mu}f_{1} & =\frac{1}{2}f_{1}\partial_{\mu}\left(H-G\right)\Rightarrow\label{df12}\\
f_{1} & =\lambda e^{\frac{1}{2}\left(H-G\right)}.\nonumber \end{align}

We now consider \eqref{kilS3} and \eqref{ckilS3} again but this
time we multiply by $-\imath\bar{\varepsilon}\gamma_{\mu}\gamma_{7}$
and $-i\gamma_{\mu}\gamma_{7}\varepsilon$ respectively and add them.
The result of the operation reads\[
\alpha L_{\mu}e^{-\frac{1}{2}\left(G+H\right)}+\frac{1}{2}f_{1}\partial_{\mu}\left(G+H\right)+\frac{1}{2}\star\left(\Omega\wedge F\right)_{\mu}e^{-\frac{3}{2}\left(G+H\right)}=0.\]
 Using \eqref{df1} and \ref{df12} we obtain the relation\begin{equation}
\partial_{\mu}e^{H}=-\frac{\alpha}{\lambda}L_{\mu}.\label{dHL}\end{equation}
 Multiplying \ref{kilS1} and \ref{ckilS1} by $\imath\bar{\varepsilon}\gamma_{\mu}\gamma_{7}$
respectively $\imath\gamma_{\mu}\gamma_{7}\varepsilon$ and adding
the resulting equations we obtain\[
-ne^{-\frac{1}{2}\left(H-G\right)}L_{\mu}-\frac{1}{4}e^{\frac{1}{2}\left(H-G\right)}\star\Omega_{\mu\lambda\nu}\mathcal{F}^{\lambda\nu}+\frac{1}{2}f_{2}\partial_{\mu}\left(H-G\right)-\frac{1}{2}e^{-\frac{3}{2}\left(G+H\right)}F_{\mu}^{\:\lambda}K_{\lambda}=0.\]
 We now use \ref{df2} and \ref{df22} to obtain\begin{equation}
\partial_{\mu}e^{H}=\frac{n}{\kappa}L_{\mu}+\frac{1}{4}e^{\frac{1}{2}\left(H-G\right)}\star\Omega_{\mu\lambda\nu}\mathcal{F}^{\lambda\nu}\label{H}\end{equation}
 which in combination with \eqref{dHL} gives the constraint\[
\star\Omega_{\mu\lambda\nu}\mathcal{F}^{\lambda\nu}=-4e^{-H+G}\left(\frac{n}{\kappa}+\frac{\alpha}{\lambda}\right)L_{\mu}.\]
 From equation \eqref{kil1} one can also derive the constraints\begin{align*}
-\alpha e^{-\frac{1}{2}\left(G+H\right)}K_{\mu}+V_{\mu}^{\:\:\lambda}\partial_{\lambda}H-\frac{1}{2}e^{\frac{1}{2}\left(H-G\right)}\mathcal{F}_{\mu\lambda}L^{\lambda} & =0\\
ne^{-\frac{1}{2}\left(H-G\right)}K_{\mu}+Y_{\mu}^{\:\:\lambda}\partial_{\lambda}H+\frac{1}{4}e^{\frac{1}{2}\left(H-G\right)}\Omega_{\mu\lambda\nu}\mathcal{F}^{\lambda\nu} & =0\end{align*}

\subsubsection*{Rank Three Identities}

We now multiply \eqref{kil1} by $\bar{\varepsilon}\gamma_{\mu\nu\kappa}$,
$\bar{\varepsilon}\gamma_{\mu\nu\kappa}\gamma_{7}$ and take the real
and imaginary part separately\begin{align}
\alpha e^{-\frac{1}{2}\left(G+H\right)}\Omega_{\mu\nu\kappa}+\frac{1}{2}\epsilon_{\mu\nu\kappa\rho\alpha\beta}Y^{\alpha\beta}\partial^{\rho}H+\frac{3}{2}e^{\frac{1}{2}\left(H-G\right)}\mathcal{F}_{\left[\kappa\right.}^{\quad\beta}\star\Omega_{\left.\mu\nu\right]\beta} & =0\nonumber \\
ne^{-\frac{1}{2}\left(H-G\right)}\star\Omega_{\mu\nu\kappa}+3\partial_{\left[\kappa\right.}HV_{\left.\mu\nu\right]}+\frac{3}{2}e^{\frac{1}{2}\left(H-G\right)}L_{\left[\kappa\right.}\mathcal{F}_{\left.\mu\nu\right]}+\frac{1}{4}e^{\frac{1}{2}\left(H-G\right)}\epsilon_{\kappa\mu\nu\alpha\beta\gamma}K^{\gamma}\mathcal{F}^{\alpha\beta} & =0\label{Omega_wedge_L}\\
\alpha e^{-\frac{1}{2}\left(G+H\right)}\star\Omega_{\mu\nu\kappa}+3\partial_{\left[\kappa\right.}HY_{\left.\mu\nu\right]}+\frac{3}{2}e^{\frac{1}{2}\left(H-G\right)}\mathcal{F}_{\left[\kappa\right.}^{\quad\beta}\Omega_{\left.\mu\nu\right]\beta} & =0\label{alpha_star_Omega}\\
ne^{-\frac{1}{2}\left(H-G\right)}\Omega_{\mu\nu\kappa}+\frac{1}{2}\epsilon_{\mu\nu\kappa\rho\alpha\beta}V^{\alpha\beta}\partial^{\rho}H+\frac{3}{2}e^{\frac{1}{2}\left(G+H\right)}K_{\left[\kappa\right.}\mathcal{F}_{\left.\mu\nu\right]}+\frac{1}{4}e^{\frac{1}{2}\left(H-G\right)}\epsilon_{\kappa\mu\nu\alpha\beta\gamma}L^{\gamma}\mathcal{F}^{\alpha\beta} & =0\label{Omega_n}\end{align}

\subsubsection*{Kahler Structure}

At this point we would like to recognize a Kahler structure in the
construction at hand. For this reason we consider \eqref{Fierz} with
$\bar{\psi}_{1}=\bar{\varepsilon}_{\pm}\gamma^{\mu\nu}\gamma^{\alpha}$,
$\psi_{2}=\varepsilon_{\pm}$, $\bar{\psi}_{3}=\bar{\varepsilon}_{\pm}$
and $\psi_{4}=\gamma_{\alpha}\gamma^{\gamma\delta}\varepsilon_{\pm}$
which gives\begin{align*}
4\,\bar{\varepsilon}_{\pm}\gamma^{\mu\nu}\gamma^{\alpha}\varepsilon_{\pm}\bar{\varepsilon}_{\pm}\gamma_{\alpha}\gamma^{\gamma\delta}\varepsilon_{\pm} & =\bar{\varepsilon}_{\pm}\gamma^{\mu\nu}\gamma^{\alpha}\gamma_{\rho}\gamma_{\alpha}\gamma^{\gamma\delta}\varepsilon_{\pm}\bar{\varepsilon}_{\pm}\gamma^{\rho}\varepsilon_{\pm}-\frac{1}{12}\bar{\varepsilon}_{\pm}\gamma^{\mu\nu}\gamma^{\alpha}\gamma_{\rho\sigma\tau}\gamma_{\alpha}\gamma^{\gamma\delta}\varepsilon_{\pm}\bar{\varepsilon}_{\pm}\gamma^{\rho\sigma\tau}\varepsilon_{\pm}\\
 & =-4\bar{\varepsilon}_{\pm}\gamma^{\mu\nu}\gamma_{\rho}\gamma^{\gamma\delta}\varepsilon_{\pm}\bar{\varepsilon}_{\pm}\gamma^{\rho}\varepsilon_{\pm}.\end{align*}
 Using the identities\begin{align}
\gamma_{\mu\nu}\gamma_{\alpha} & =g_{\alpha\nu}\gamma_{\mu}-g_{\alpha\mu}\gamma_{\nu}+\gamma_{\alpha\mu\nu}\label{id1}\\
\gamma_{\alpha}\gamma_{\gamma\delta} & =g_{\alpha\gamma}\gamma_{\delta}-g_{\alpha\delta}\gamma_{\gamma}+\gamma_{\alpha\gamma\delta}\label{id2}\\
\gamma_{\mu\nu}\gamma_{\alpha}\gamma_{\gamma\delta} & =\gamma_{\alpha\gamma\delta\mu\nu}-6g_{\alpha\pi}g_{\gamma\rho}g_{\delta\sigma}\delta_{\:\mu\nu}^{\left[\pi\rho\right.}\gamma^{\left.\sigma\right]}+4g_{\alpha\pi}g_{\mu\rho}g_{\nu\sigma}\delta_{\:\gamma\delta}^{\pi\left[\rho\right.}\gamma^{\left.\sigma\right]}\nonumber \\
 & \quad+6g_{\alpha\pi}g_{\gamma\rho}g_{\delta\sigma}\delta_{\left[\nu\right.}^{\left[\pi\right.}\gamma_{\quad\:\left.\mu\right]}^{\left.\rho\sigma\right]}+2g_{\alpha\left[\gamma\right.}\gamma_{\left.\delta\right]\mu\nu}\nonumber \\
\gamma^{\alpha}\gamma_{\rho\sigma\tau}\gamma_{\alpha} & =0\nonumber \end{align}
 we draw the conclusion\begin{align}
I_{\: b}^{a}I_{\: c}^{b} & =-\delta_{c}^{a}\label{II}\\
J_{\: b}^{a}J_{\: c}^{b} & =-\delta_{c}^{a}.\nonumber \end{align}
 Using \eqref{l+l+}, \eqref{l-l-}, \eqref{II}, \eqref{ZZ}, \eqref{q+q-}
and the definitions \eqref{Z+}-\eqref{l-} we have that\begin{align}
L^{2} & =-K^{2}=f_{1}^{2}+f_{2}^{2}\label{L2K2}\\
K\cdot L & =0.\label{KL}\end{align}

We now look at the Fierz identity involving $\bar{\psi}_{1}=\bar{\varepsilon}_{+}\gamma^{\gamma\delta}\gamma_{\mu}$,
$\psi_{2}=\varepsilon_{+}$, $\bar{\psi}_{3}=\bar{\varepsilon}_{-}$
and $\psi_{4}=\gamma_{\nu}\gamma_{\gamma\delta}\varepsilon_{-}$ which
gives after antisymmetrization in $\mu$ and $\nu$\begin{align}
4\;\bar{\varepsilon}_{+}\gamma_{\quad\left[\mu\right.}^{\gamma\delta}\varepsilon_{+}\;\bar{\varepsilon}_{-}\gamma_{\left.\nu\right]\gamma\delta}\varepsilon_{-} & =-12\bar{\varepsilon}_{+}\gamma_{\mu\nu}\varepsilon_{-}\;\bar{\varepsilon}_{-}\varepsilon_{+}+12\bar{\varepsilon}_{+}\varepsilon_{-}\;\bar{\varepsilon}_{-}\gamma_{\mu\nu}\varepsilon_{+}\nonumber \\
 & \quad-2\bar{\varepsilon}_{+}\gamma_{\mu\nu\rho\sigma}\varepsilon_{-}\;\bar{\varepsilon}_{-}\gamma^{\rho\sigma}\varepsilon_{+},\label{double}\end{align}
 considering now the Fierz identity for $\bar{\psi}_{1}=\bar{\varepsilon}_{+}\gamma_{\nu}$,
$\psi_{2}=\varepsilon_{+}$, $\bar{\psi}_{3}=\bar{\varepsilon}_{-}$
and $\psi_{4}=\gamma_{\mu}\varepsilon_{-}$we have\[
\bar{\varepsilon}_{+}\gamma_{\mu\nu\rho\sigma}\varepsilon_{-}\;\bar{\varepsilon}_{-}\gamma^{\rho\sigma}\varepsilon_{+},=8\bar{\varepsilon}_{+}\gamma_{\left[\nu\right.}\varepsilon_{+}\bar{\varepsilon}_{-}\gamma_{\left.\mu\right]}\varepsilon_{-}+2\bar{\varepsilon}_{+}\gamma_{\mu\nu}\varepsilon_{-}\;\bar{\varepsilon}_{-}\varepsilon_{+}-2\bar{\varepsilon}_{+}\varepsilon_{-}\;\bar{\varepsilon}_{-}\gamma_{\mu\nu}\varepsilon_{+}.\]
 Finally \eqref{double} takes the form\begin{equation}
\bar{\varepsilon}_{+}\gamma_{\quad\left[\mu\right.}^{\gamma\delta}\varepsilon_{+}\;\bar{\varepsilon}_{-}\gamma_{\left.\nu\right]\gamma\delta}\varepsilon_{-}=-4\bar{\varepsilon}_{+}\gamma_{\mu\nu}\varepsilon_{-}\;\bar{\varepsilon}_{-}\varepsilon_{+}+4\bar{\varepsilon}_{+}\varepsilon_{-}\;\bar{\varepsilon}_{-}\gamma_{\mu\nu}\varepsilon_{+}-4\bar{\varepsilon}_{+}\gamma_{\left[\nu\right.}\varepsilon_{+}\bar{\varepsilon}_{-}\gamma_{\left.\mu\right]}\varepsilon_{-}.\label{opa}\end{equation}
 It will be also useful to consider the Fierz identities with the
choice $\bar{\psi}_{1}=\bar{\varepsilon}_{-}\gamma^{\mu}$, $\psi_{2}=\varepsilon_{-}$,
$\bar{\psi}_{3}=\bar{\varepsilon}_{-}$ and $\psi_{4}=\gamma_{\mu}\gamma_{\nu}\varepsilon_{+}$
which gives the relation\begin{align}
\bar{\varepsilon}_{-}\gamma^{\mu}\varepsilon_{-}\bar{\varepsilon}_{-}\gamma_{\mu}\gamma_{\nu}\varepsilon_{+} & =0\Rightarrow\nonumber \\
\bar{\varepsilon}_{-}\gamma^{\mu}\varepsilon_{-}\bar{\varepsilon}_{-}\gamma_{\mu\nu}\varepsilon_{+} & =-\bar{\varepsilon}_{-}\gamma_{\nu}\varepsilon_{-}\bar{\varepsilon}_{-}\varepsilon_{+}.\label{l+U+}\end{align}
 In a similar way one can derive\begin{equation}
\bar{\varepsilon}_{+}\gamma^{\mu}\varepsilon_{+}\bar{\varepsilon}_{+}\gamma_{\mu\nu}\varepsilon_{-}=-\bar{\varepsilon}_{+}\gamma_{\nu}\varepsilon_{+}\bar{\varepsilon}_{+}\varepsilon_{-}.\label{l-U-}\end{equation}

Contracting \eqref{Omega_n} with $K^{\mu}L^{\nu}$ and using \eqref{dHL},
\eqref{KL} and \eqref{condition} we obtain\begin{equation}
L^{\mu}\mathcal{F}_{\mu\nu}=0.\label{LW}\end{equation}
 Contracting \eqref{Omega_n} with $L^{\nu}$we obtain the equation\[
n\Omega_{\mu\lambda\nu}L^{\nu}=0\]
 which after combining with \eqref{H} gives\begin{align}
n\left(i_{l^{+}}q^{-}-i_{l^{-}}q^{+}\right) & =0\label{lqlq}\end{align}
 where we used \[
i_{l^{\pm}}q^{\pm}=0.\]
 From \eqref{lqlq}, \eqref{Idef} and \eqref{Jdef} we conclude that
\[
n\left(I-J\right)=0.\]
 From now on we assume that $n\neq0$ and we conclude that\[
I=J.\]

Using equations \eqref{kilS3}, \eqref{kilS1}, \eqref{ckilS3} and
\eqref{ckilS1} one can show that \begin{align*}
\bar{\varepsilon}\left(\gamma_{\mu\nu\lambda\kappa}N+N\gamma_{\kappa\mu\nu\lambda}\right)\gamma_{7}\varepsilon & =\frac{1}{2}\left(\star\Omega\wedge dG\right)_{\kappa\mu\nu\lambda}-\frac{1}{2}e^{\frac{1}{2}\left(H-G\right)}\mathcal{F}_{\:\:\left[\kappa\right.}^{\pi}\varepsilon_{\left.\mu\nu\lambda\right]\pi\alpha\beta}Y^{\alpha\beta}.\end{align*}
 Combining the above equation with \eqref{d*omega} we have that\begin{equation}
d\star\Omega=\star\Omega\wedge dG-2e^{\frac{1}{2}\left(H-G\right)}\mathcal{F}_{\:\:\left[\kappa\right.}^{\pi}\varepsilon_{\left.\mu\nu\lambda\right]\pi\alpha\beta}Y^{\alpha\beta}.\label{dif_star_Omega}\end{equation}
 One can easily check that because of \eqref{condition}, \eqref{LW}
the second term vanishes. Equation \eqref{dif_star_Omega} show us
then that for the form\begin{align*}
U & =e^{G}\star\Omega\Rightarrow\\
dU & =e^{G}\left(dG\wedge\star\Omega+d\star\Omega\right)\\
 & =e^{G}\left(dG\wedge\star\Omega+\star\Omega\wedge dG\right)=0.\end{align*}
 We know that \[
U=e^{G}L\wedge I\]
 and since $L$ is closed we have that\begin{equation}
L\wedge d\left(e^{G}I\right)=0.\label{LwedgedJ}\end{equation}
 Rescaling the vielbein in \eqref{metric} as\[
e_{\mu}^{a}=e^{-\frac{1}{2}G-\frac{1}{2}H}\tilde{e}_{\mu}^{a}\]
 we define the metric \begin{equation}
h_{\mu\nu}=\delta_{ab}\tilde{e}_{\mu}^{a}\tilde{e}_{\nu}^{b}\label{hmn}\end{equation}
 and the two-form \begin{equation}
\mathcal{J}=e^{G+H}I.\label{J}\end{equation}
 At this point we would like to fix our gauge so that\begin{align*}
e^{+} & =X\left(dt+C_{m}dx^{m}\right)+Bdy\\
e^{-} & =-X\left(dt+C_{m}dx^{m}\right)+Bdy\\
K & =-Xdt-XC\\
L & =\gamma dy\end{align*}
 and from equations \eqref{L2K2} and \eqref{KL} we draw the conclusion
that\[
X=B^{-1}=h^{-2}=f_{1}^{2}+f_{2}^{2}.\]
 At this point the ten dimensional metric has the form\[
ds^{2}=-\frac{1}{h^{2}}\left(dt+C\right)^{2}+\gamma^{2}h^{2}dy^{2}+e^{-G-H}h_{mn}dx^{m}dx^{n}-\frac{\gamma\alpha}{\lambda}ye^{G}d\hat{\Omega}_{3}^{2}-\frac{\gamma\alpha}{\lambda}ye^{-G}d\left(\psi+A\right)^{2},\qquad m=1,\ldots,4\]
 where we used \eqref{H} to fix\[
e^{H}=-\frac{\gamma\alpha}{\lambda}y.\]
 We also know that the Euclidean space with metric\[
ds_{4}^{2}=h_{mn}dx^{m}dx^{n}\]
 is Kahler with complex structure defined by \eqref{J}. In order
to prove that $\mathcal{J}$ is closed we split the exterior derivative
as\[
d=\tilde{d}+d_{y}+d_{t}\]
 where $\tilde{d}$ only takes into account differentiation with respect
to the coordinates $x^{m}$. Then from \eqref{LwedgedJ} \begin{align}
L\wedge d\mathcal{J} & =0\Rightarrow\nonumber \\
\tilde{d}\mathcal{J} & =0\label{dJ}\end{align}
 and from \eqref{II} we have\[
\mathcal{J}_{\quad p}^{m}\mathcal{J}_{\quad n}^{p}=-\delta_{n}^{m}\]
 where in this equation we raise and lower indices using the metric
\eqref{hmn}.

From equation \eqref{opa} we derive the constrain\begin{equation}
f_{1}V+f_{2}Y=L\wedge K.\label{fVfY}\end{equation}
 And from equations \eqref{opa}, \eqref{l+U+} and \eqref{l-U-}
we can derive the equations \begin{align*}
i_{L}V & =f_{1}K\\
i_{K}V & =-f_{1}L\\
i_{L}Y & =f_{2}K\\
i_{K}Y & =-f_{2}L.\end{align*}
 We now use these relations and contract equations \eqref{Omega_wedge_L}
and \eqref{alpha_star_Omega} with $L^{\kappa}$ to derive\begin{align}
V= & \frac{\lambda n}{\alpha}e^{\frac{1}{2}\left(H+G\right)}I+\frac{\lambda}{2\alpha}e^{H}e^{\frac{1}{2}\left(H-G\right)}\left(\mathcal{F}-\frac{1}{f_{1}^{2}+f_{2}^{2}}i_{L}i_{K}\star_{6}\mathcal{F}\right)+\frac{f_{1}}{f_{1}^{2}+f_{2}^{2}}\, L\wedge K\label{VI}\\
Y= & \lambda e^{\frac{1}{2}\left(H-G\right)}I+\frac{f_{2}}{f_{1}^{2}+f_{2}^{2}}\, L\wedge K.\nonumber \end{align}
 In order to satisfy \eqref{fVfY} we find that the following relations
should be satisfied\begin{equation}
\mathcal{F}-\frac{1}{f_{1}^{2}+f_{2}^{2}}i_{L}i_{K}\star_{6}\mathcal{F}=-2e^{G-H}\left(n+\frac{\kappa\alpha}{\lambda}\right)I.\label{w_constrain}\end{equation}

At this point we may use \eqref{VI} and \eqref{w_constrain} to express\begin{equation}
I=-\frac{1}{\kappa}e^{-\frac{1}{2}\left(H+G\right)}V+\frac{1}{\kappa}\frac{f_{1}e^{-\frac{1}{2}\left(G+H\right)}}{f_{1}^{2}+f_{2}^{2}}L\wedge K.\label{IequalsV}\end{equation}
 One can calculate the derivative of $V$ using equation \eqref{kileq2}
and \eqref{kil2},\begin{align*}
dV_{\lambda\mu\nu} & =e^{\frac{1}{2}\left(H-G\right)}L\wedge\mathcal{F}-\frac{1}{2}\frac{\kappa\alpha}{\lambda}e^{-\frac{1}{2}\left(H-G\right)}L\wedge I-\frac{1}{2}V\wedge dG\Rightarrow\\
d\left(e^{\frac{1}{2}G}V\right) & =e^{\frac{1}{2}H}L\wedge\mathcal{F}-\frac{1}{2}\frac{\kappa\alpha}{\lambda}e^{-\frac{1}{2}\left(H-2G\right)}L\wedge I.\end{align*}
 The above equation in combination with \eqref{IequalsV} and the
fact that\[
dK=\frac{f_{2}^{2}-f_{1}^{2}}{f_{2}^{2}+f_{1}^{2}}dG\wedge K+dH\wedge K-\left(f_{2}^{2}+f_{1}^{2}\right)dC\]
 leads to the differential \begin{align*}
d\left(e^{H+G}I\right) & =e^{H}L\wedge\left(\frac{\lambda}{\kappa}dC-\frac{1}{\kappa}\mathcal{F}\right)\\
d\mathcal{J} & =e^{H}L\wedge\left(\frac{\lambda}{\kappa}dC-\frac{1}{\kappa}\mathcal{F}\right)\end{align*}
 We split the exterior derivative as\[
d=\tilde{d}+d_{y}+d_{t}\]
 where $\tilde{d}$ only takes into account differentiation with respect
to the coordinates $x^{m}$. From the above we recover the fact that
$\mathcal{J}$ is closed and we find also an equation for the $y$
dependence of this function\begin{equation}
\partial_{y}\mathcal{J}=e^{H}\frac{\gamma}{\kappa}\left(\lambda\,\tilde{d}C-\mathcal{F}\right).\label{dyJ_dC_W}\end{equation}

\subsubsection*{Eliminating the field strength}

We can use equation \eqref{dOmega_equals_F} to solve for the four
form\begin{align*}
4f_{1}e^{-\frac{1}{2}\left(H-G\right)}\tilde{F}_{\left(4\right)} & =d\Omega-\, e^{\frac{1}{2}\left(H-G\right)}Y\wedge\mathcal{F}\\
 & =-\frac{2f_{1}^{2}}{f_{1}^{2}+f_{2}^{2}}\, dG\wedge K\wedge I+f_{1}e^{-\frac{1}{2}\left(G+H\right)}\, L\wedge K\wedge dC-\left(f_{2}^{2}+f_{1}^{2}\right)\: dC\wedge I\\
 & \quad+\lambda e^{H-G}I\wedge\mathcal{F}-e^{\frac{1}{2}\left(H-G\right)}\frac{f_{1}^{2}}{f_{2}}\frac{1}{f_{1}^{2}+f_{2}^{2}}L\wedge K\wedge\mathcal{F}\end{align*}
 From the Bianchi identity we have the constrain that $W$ has to
satisfy\begin{equation}
\mathcal{F}\wedge d\left(e^{\frac{1}{2}\left(H-G\right)}Y\right)=0.\label{W_Y_constrain}\end{equation}
 Contracting equation \eqref{df1} with $L^{\mu}$ we obtain the equation\begin{align*}
L^{\mu}\partial_{\mu}f_{1}^{2} & =-\frac{1}{4}d\Omega_{\mu\rho\sigma\tau}L^{\mu}K^{\rho}I^{\sigma\tau}\\
\frac{1}{4}e^{\frac{1}{2}\left(G+H\right)}\left(f_{1}^{2}+f_{2}^{2}\right)\,\mathcal{J}^{ij}\tilde{d}C_{ij} & =\gamma f_{1}\partial_{y}G+\gamma f_{1}e^{-H}\partial_{y}e^{H}-e^{-\frac{1}{2}\left(H-G\right)}\frac{f_{1}}{f_{2}}\left(n+\frac{\kappa\alpha}{\lambda}\right)\end{align*}
 where we used the constrain \eqref{w_constrain} and in the last
line the raising of indices is done with the Kahler metric. Contracting
equation \eqref{dyJ_dC_W} with $\mathcal{J}$ we obtain\begin{align*}
\mathcal{J}^{ij}\partial_{y}\mathcal{J}_{ij} & =\frac{\gamma\lambda}{\kappa}e^{H}\mathcal{J}^{ij}dC_{ij}-\frac{\gamma}{\kappa}e^{H}\mathcal{J}^{ij}\mathcal{F}_{ij}\end{align*}
 where the contraction is done using the rescaled metric \eqref{hmn}.
At this point we find it convenient to set\begin{align*}
\kappa & =\lambda=1\\
\gamma^{2} & =1\\
\gamma\alpha & =-1\Rightarrow e^{H}=y.\end{align*}
 Continuing our analysis we have that\begin{align*}
\partial_{y}\ln\det\mathcal{G} & =2\frac{e^{-G}}{e^{G}+e^{-G}}\partial_{y}G+2\gamma\frac{\gamma-n-\alpha}{y\left(1+e^{2G}\right)}+\frac{2\gamma}{y}\left(n+\alpha\right)\end{align*}
 where\[
\mathcal{G}=\left|\begin{array}{cc}
\partial_{z}\partial_{\bar{z}}K & \partial_{w}\partial_{\bar{z}}K\\
\partial_{z}\partial_{\bar{w}}K & \partial_{w}\partial_{\bar{w}}K\end{array}\right|\]
 and $K$ represents the Kahler potential for the four dimensional
base manifold.

Defining the field\[
Z=\frac{1}{2}\tanh G\]
 the previous equations may be written as\begin{equation}
\partial_{y}\ln\det\mathcal{G}=\partial_{y}\ln\left(Z+\frac{1}{2}\right)+\frac{2\gamma}{y}\left(-Z+\frac{1}{2}\right)\left(\gamma-n-\alpha\right)+\frac{2\gamma}{y}\left(n+\alpha\right).\label{y_scalar_eq}\end{equation}
 We are now turning our attention to the component\begin{align*}
F_{tx} & =\frac{e^{2G+H}}{4!}\epsilon_{tx}^{\quad\pi\rho\sigma\tau}F_{\pi\rho\sigma\tau}\\
 & =-\frac{1}{4}\frac{e^{\frac{3}{2}\left(G+H\right)}}{f_{1}}\left(f_{1}^{2}+f_{2}^{2}\right)^{2}\partial_{y}C_{x^{1}}I_{x}^{\; x^{1}}.\end{align*}
 Using equation \eqref{df2} we have the constrain\begin{align*}
I_{x}^{\; x^{1}}\partial_{y}C_{x^{1}} & =-\frac{2f_{1}f_{2}}{\left(f_{1}^{2}+f_{2}^{2}\right)^{2}}\partial_{x}G\\
\partial_{y}C_{x^{1}} & =\frac{1}{y}I_{x}^{\; x^{1}}\partial_{x^{1}}Z.\end{align*}
 The consistency condition $d^{2}C=0$ leads to the following expression
between the scalar $Z$ and the Kahler form $\mathcal{J}$\begin{equation}
y\partial_{y}\left(\frac{\partial_{y}\mathcal{J}}{y}\right)+\tilde{d}\left(\mathcal{J}\cdot\tilde{d}Z\right)=0.\label{LLM_type}\end{equation}

\subsubsection*{The Ricci form of the Kahler base space}

We would like to construct the holomorphic 2-form $\mathcal{B}$ for
the 4 dimensional Kahler base space such that\[
\mathcal{B}\wedge\bar{\mathcal{B}}=\mathcal{J}\wedge\mathcal{J}.\]
 For this reason we consider the bilinear\[
\mathcal{V}_{\mu\nu}=\varepsilon^{T}\gamma_{\mu\nu}\gamma_{7}\varepsilon.\]
 Our convention for the gamma matrices are such that\begin{align*}
\gamma_{\mu}^{\star} & =\gamma_{0}\gamma_{\mu}\gamma_{0}\\
\gamma_{\mu}^{T} & =-\gamma_{\mu}.\end{align*}
 Before we move on with our study we note the projection condition
that the spinor has to satisfy as a result of the Fierz identities\begin{equation}
\gamma_{0}\gamma_{7}\gamma_{\bar{y}}\varepsilon=\alpha\varepsilon.\label{projection}\end{equation}
 Fierzing for $\psi_{4}=-\gamma_{0}\gamma_{\kappa\lambda}\gamma_{7}\gamma_{0}\varepsilon^{\star}$,
$\psi_{2}=\varepsilon$, $\psi_{3}=\gamma_{0}\varepsilon$ and $\psi_{1}=-\gamma_{7}\gamma_{\mu\nu}\gamma_{0}\varepsilon^{\star}$
we find that\[
\mathcal{V}\wedge\bar{\mathcal{V}}=e^{-2\left(H+G\right)}\left(f_{1}^{2}+f_{2}^{2}\right)\,\mathcal{J}\wedge\mathcal{J}.\]
 We consider one more time the Fierz identity for $\psi_{4}=-\gamma_{0}\gamma_{\mu\nu}\gamma_{7}\gamma_{0}\varepsilon^{\star}$,
$\psi_{2}=\varepsilon$, $\psi_{3}=\gamma_{0}\varepsilon$ and $\psi_{1}=-\gamma_{\pi\rho\sigma}\varepsilon$
with which we may prove that\[
i_{K}\Omega\wedge\mathcal{V}=e^{-G-H}\mathcal{J}\wedge\mathcal{V}=0\]
 which proves that $\mathcal{V}$ is holomorphic. At this point we
note that we used \eqref{projection} to show that\[
i_{K}\mathcal{V}=\sqrt{f_{1}^{2}+f_{2}^{2}}\varepsilon^{T}\gamma_{\nu}\gamma_{0}\gamma_{7}\varepsilon=\alpha\sqrt{f_{1}^{2}+f_{2}^{2}}\varepsilon^{T}\gamma_{\nu}\gamma_{\bar{y}}\varepsilon=0\]
 as a consequence of the antisymmetry of the antisymmetry of the gamma
matrices. We may also use \eqref{kilS3} to show that\[
\varepsilon^{T}\varepsilon=0.\]
 From the above analysis we see that\begin{equation}
\mathcal{V}=e^{-\left(H+G\right)}\sqrt{f_{1}^{2}+f_{2}^{2}}\,\mathcal{B}.\label{mathcal_V_equals_mathcal_B}\end{equation}
 We may now use \eqref{kileq2} to show that\begin{align*}
d\mathcal{V}_{\rho\mu\nu} & =inA\wedge\mathcal{V}_{\rho\mu\nu}-\frac{3}{2}e^{\frac{1}{2}\left(H-G\right)}\varepsilon^{T}\gamma_{\left[\mu\nu\right.}^{\quad\lambda}\varepsilon\mathcal{F}_{\left.\lambda\rho\right]}-\imath\varepsilon^{T}\left(\gamma_{\rho\mu\nu}N-N\gamma_{\rho\mu\nu}\right)\gamma_{7}\varepsilon\\
 & =inA\wedge\mathcal{V}_{\rho\mu\nu}-\frac{3}{2}e^{\frac{1}{2}\left(H-G\right)}\varepsilon^{T}\gamma_{\left[\mu\nu\right.}^{\quad\lambda}\varepsilon\mathcal{F}_{\left.\lambda\rho\right]}-\alpha e^{-\frac{1}{2}\left(H+G\right)}\varepsilon^{T}\gamma_{\rho\mu\nu}\gamma_{7}\varepsilon-\frac{1}{2}\mathcal{V}\wedge d\left(H+G\right)_{\rho\mu\nu}.\end{align*}
 We are mostly interested in the case where the indices $\rho,\,\mu,\,\nu$
belong to the four dimensional base space. For this reason one can
write\begin{align}
d\mathcal{V}_{rmn} & =inA\wedge\mathcal{V}_{rmn}+\frac{3}{2}e^{\frac{1}{2}\left(H-G\right)}\sqrt{f_{1}^{2}+f_{2}^{2}}C_{\left[m\right.}\varepsilon^{T}\gamma_{0}\gamma_{n}^{\quad l}\varepsilon\mathcal{F}_{\left.lr\right]}\nonumber \\
 & \quad+\alpha e^{-\frac{1}{2}\left(H+G\right)}\sqrt{f_{1}^{2}+f_{2}^{2}}C_{\left[r\right.}\varepsilon^{T}\gamma_{0}\gamma_{\left.mn\right]}\gamma_{7}\varepsilon-\frac{1}{2}\mathcal{V}\wedge d\left(H+G\right)_{rmn}.\label{dmathcal_V}\end{align}
 Where we used the fact that because of \eqref{projection}\[
\bar{\varepsilon}\gamma_{\bar{r}\bar{m}\bar{n}}\varepsilon=0.\]
 There are three useful relations that we can derive from the projector
\eqref{kil1}\begin{align*}
-\imath ne^{-\frac{1}{2}\left(H-G\right)}\mathcal{V}_{mn}-\frac{1}{2}e^{\frac{1}{2}\left(H-G\right)}\mathcal{V}_{\left[ma\right.}\mathcal{F}_{\:\left.n\right]}^{a}-\varepsilon^{T}\gamma_{mn}^{\quad l}\varepsilon\,\partial_{l}H & =0\\
\alpha e^{-\frac{1}{2}\left(H+G\right)}\varepsilon^{T}\gamma_{mnl}\gamma_{7}\varepsilon-\imath ne^{-\frac{1}{2}\left(H-G\right)}\varepsilon^{T}\gamma_{mnl}\varepsilon-\frac{3}{2}e^{\frac{1}{2}\left(H-G\right)}\varepsilon^{T}\gamma_{\left[mn\right.a}\varepsilon\,\mathcal{F}_{\:\left.l\right]}^{a} & =0\\
\alpha e^{-\frac{1}{2}\left(H+G\right)}\mathcal{V}_{mn}+\varepsilon^{T}\gamma_{mnl}\gamma_{7}\varepsilon\,\partial^{l}H & =0.\end{align*}
 At this points it is useful to note that one can easily prove that
$\mathcal{F}$ has to be a $\left(1,1\right)$ form and because $\mathcal{V}$
is a holomorphic form we may write\[
\mathcal{V}_{\left[ma\right.}\mathcal{F}_{\:\left.n\right]}^{a}=\frac{1}{2}\imath I^{ab}\mathcal{F}_{ab}\,\mathcal{V}_{mn}=-2\imath e^{G-H}\left(n+\alpha\right)\,\mathcal{V}_{mn}.\]
 From the first equation we have that\[
e^{-\frac{1}{2}\left(H+G\right)}\sqrt{f_{1}^{2}+f_{2}^{2}}\,\varepsilon^{T}\gamma_{0}\gamma_{mn}\gamma_{7}\varepsilon=\imath\alpha\mathcal{V}_{mn}.\]
 Using the second and third lines we may write\[
\frac{3}{2}e^{\frac{1}{2}\left(H-G\right)}\sqrt{f_{1}^{2}+f_{2}^{2}}C_{\left[m\right.}\varepsilon^{T}\gamma_{0}\gamma_{n}^{\quad l}\varepsilon\mathcal{F}_{\left.lr\right]}=\imath\left(n+\alpha\right)C_{\left[m\right.}\mathcal{V}_{\left.mn\right]}.\]
 Substituting the above two equations in \eqref{dmathcal_V} we obtain\[
\tilde{d}\mathcal{V}=\imath\,\left[nA+\left(2\alpha+n\right)C-\frac{\imath}{2}\tilde{d}G\right]\wedge\mathcal{V}.\]
 We may now use \eqref{mathcal_V_equals_mathcal_B} to calculate\[
\tilde{d}\mathcal{B}=\left[\imath nA+\imath\left(2\alpha+n\right)C+\frac{1}{2}\tilde{d}\ln\left(Z+\frac{1}{2}\right)\right]\wedge\mathcal{B}.\]
 From the above we have that the Ricci form of the Kahler manifold
is given by\begin{align}
\mathcal{R} & =-n\mathcal{F}-\left(2\alpha+n\right)\tilde{d}C+\frac{1}{2}d\left(\mathcal{J}\cdot d\ln\left(Z+\frac{1}{2}\right)\right)\nonumber \\
 & =-2\left(\alpha+n\right)\mathcal{F}+\frac{\alpha}{y}\left(2\alpha+n\right)\partial_{y}\mathcal{J}+\frac{1}{2}d\left(\mathcal{J}\cdot d\ln\left(Z+\frac{1}{2}\right)\right)\label{Ricci_Form}\end{align}
 which is compatible with equation \eqref{y_scalar_eq}.

We may now use the constrain \eqref{WVconstrain} to write\[
\mathcal{F}=\tilde{d}\left(\mathcal{J}\cdot\tilde{d}\Phi\right).\]
 Finally from \eqref{y_scalar_eq} and \eqref{Ricci_Form} we have
that\begin{align}
\ln\left|\begin{array}{cc}
\partial_{z}\partial_{\bar{z}}K & \partial_{w}\partial_{\bar{z}}K\\
\partial_{z}\partial_{\bar{w}}K & \partial_{w}\partial_{\bar{w}}K\end{array}\right| & =\ln\left(-y\partial_{y}\left(\frac{\partial_{y}K}{y}\right)+yF^{\prime}\left(y\right)\right)+\frac{2\alpha}{y}\left(2\alpha+n\right)\partial_{y}K+4\alpha\left(\alpha+n\right)\Phi\nonumber \\
 & \qquad-2\alpha\left(n+\alpha\right)\ln y-2\alpha\left(2\alpha+n\right)F\left(y\right)+2\alpha\left(2\alpha+n\right)\ln y\label{mong_Amp}\end{align}
 where the scalars need to satisfy the two additional constraints\begin{align*}
\partial\bar{\partial}\left[Z+y\partial_{y}\left(\frac{\partial_{y}K}{y}\right)\right] & =0\\
\partial\bar{\partial}\Phi-\star_{4}\partial\bar{\partial}\Phi & =-\frac{2}{y^{2}}\left(n+\alpha\right)\partial\bar{\partial}K\end{align*}
 and the function $F\left(y\right)$ is such that\[
Z+\frac{1}{2}=-y\partial_{y}\left(\frac{\partial_{y}K}{y}\right)+yF^{\prime}\left(y\right).\]
 We also have to satisfy \eqref{F_4_W_constrain} and \eqref{W_Y_constrain}
which come from the Bianchi identity of the ten dimensional five form
field. We can use the constrain \eqref{w_constrain} to show that\[
I\wedge\mathcal{F}=-\left(n+\alpha\right)e^{G-H}I\wedge I.\]
 We can easily check that the constrains \eqref{F_4_W_constrain}
and \eqref{W_Y_constrain} are simultaneously solved when\[
e^{-2H}\left(n+\alpha\right)\partial_{y}\ln\left(Z+\frac{1}{2}\right)\,\mathcal{J}\wedge\mathcal{J}-e^{-H}\left(n+\alpha\right)\,\partial_{y}\left(\frac{1}{y^{2}}\mathcal{J}\wedge\mathcal{J}\right)-\alpha\,\mathcal{F}\wedge\mathcal{F}+\gamma\left(\frac{1}{2}-Z\right)\,\mathcal{F}\wedge\mathcal{F}=0.\]
 Comparing with equation \eqref{y_scalar_eq} we see that we need
to have\begin{equation}
\mathcal{F}\wedge\mathcal{F}=2\left(n+2\alpha\right)\left(n+\alpha\right)\, y^{-4}\,\mathcal{J}\wedge\mathcal{J}.\label{WW_equals_JJ}\end{equation}

\newpage{} \bibliographystyle{/sw/share/texmf-dist/bibtex/bst/arXiv/utphys}
\bibliography{Bibliography}

\end{document}